# Improving brain computer interface performance by data augmentation with conditional Deep Convolutional Generative Adversarial Networks

Qiqi Zhang, Ying Liu

*Abstract*—One of the big restrictions in brain computer interface field is the very limited training samples, it is difficult to build a reliable and usable system with such limited data. Inspired by generative adversarial networks, we proposed conditional Deep Convolutional Generative Adversarial Networks (cDCGAN) method to generate more artificial EEG signal automatically for data augmentation to improve the performance of convolutional neural networks in brain computer interface field and overcome the small training dataset problems. We evaluate the proposed cDCGAN method on BCI competition dataset of motor imagery. The results show that the generated artificial EEG data from Gaussian noise can learn the features from raw EEG data and has no less than the classification accuracy of raw EEG data in the testing dataset. Also by using generated artificial data can effectively improve classification accuracy at the same model with limited training data.

## I. INTRODUCTION

Brain computer interface (BCI) can connect brain and external world by recognizing brain activities and translating it into messages or commands and this process does not depend on normal peripheral nerves and muscles [1]. So how to recognize meaningful brain activities from seemingly meaningless brain electrical signal is a big challenge for BCI field. Up to now, many methods have been proposed to improve recognition performance in many aspects including preprocessing, feature extraction, feature selection, classification and post-processing aspects [2][3]. Although many achievements have been obtained in this field and many algorithms can get state of the art performance with high classification accuracy in standard dataset or self-recorded dataset, most of the algorithms are only evaluated offline not online running [4]. This may attribute to lack of enough data to learn more possible features to make the classifier more robust and reliable, especially for the deep learning model because it need much more data than other methods [5][6]. Compared with other fields such as computer vision and speech recognition, only very limited number of training samples can be obtained in BCI field, because the subject cannot be asked to perform the same mental task every time for much long time to record brain signal [6]. And deep learning model usually has huge numbers of parameters, so the limited training datasets cannot exploit the full potential of deep learning model in BCI field [6][7]. So, we proposed a novel approach in this paper to generate more artificial brain signal data automatically to overcome this data deficiency problem and improve the performance of deep learning model. In this preliminary study, we take electroencephalography (EEG) and convolutional neural networks (CNN) for example.

Fabien Lotte first to generate artificial EEG data by mixing signal segmentation in time domain, and their offline analysis suggested this approach can significantly increase classification accuracy when training data was small [8]. However, this approach may cause inadequate high frequency noise at the boundary of two different segments. To overcome this problem, artificial EEG signal generation methods based on time-frequency representation and analogy method were proposed [4]. The prior works only considered the temporal features of EEG signal without frequency features, so the Empirical Mode Decomposition (EMD) method was proposed to consider the features in both temporal and frequency domains [9]. However, abovementioned works are all based on classic processing method with common spatial patterns and linear discriminative analysis. For burgeoning deep learning method, differential entropy feature was used to generate more EEG signal, and this method can improve the performance of deep models (LeNet and ResNet) significantly [10].

However, up to now all the methods to generate artificial EEG signal are based on the combination of the features of raw EEG signal in different trials. For our methods, we generate artificial EEG signal from probability distribution and deep learning perspective. Instead of physically combining the effective features like signal segments, time-frequency representations, intrinsic mode function or differential entropy abovementioned, we automatically generate artificial EEG signal by a trained generative model that can approximate the feature distribution of raw EEG signal during a training of the game process. This data generating method is called generative adversarial networks (GAN) and GAN-based methods have been used in computer vision like generating image from text, generating videos with scene dynamics and image to image translation [11-14]. However, to the best of our knowledge, it's the first to attempt GAN-based method to improve the performance of BCI.

In this paper, we proposed the conditional Deep Convolutional Generative Adversarial Networks (cDCGAN) method to generate more artificial EEG signal for data augmentation to improve the performance of convolutional neural networks (CNN) and overcome the small training dataset problems. For CNN model, raw EEG signal was transformed into time-frequency representation (TFR) and we used the two-dimensional kernel to learn the time-frequency features from TFR of raw EEG signal. Because CNN model has showed superiority for feature extraction in BCIs [15], we

*Research supported by the National Natural Science Foundation of China under Grant 51775041.

The authors are with Mechanical Engineering Department, Beijing Institute of Technology, Beijing, China. {zhangqiqi, biggirlliu}@bit.edu.cn

applied it for GAN to construct cDCGAN and the CNN was used for discriminative model and the inverse process of CNN was used for generative model [16]. Therefore, cDCGAN is used to generate artificial TFR of EEG signal then the inverse of wavelet transform is applied to generate waveform EEG signal.

## II. METHODS

In order to overcome small training datasets problem, we proposed a new data augmentation method to generate artificial EEG signal by conditional Deep Convolutional Generative Adversarial Networks (cDCGAN). As Fig. 1. showed, the whole procedure comprised training part and testing part similar to common processing procedure, but in our proposed method we added data augmentation in training part with limited training data to generate more artificial EEG signal and used this artificial signal as extra training dataset. And to improve the classification accuracy of EEG signal with generated artificial EEG data, the artificial data must have label information. Inspired by the performance of conditional Generative Adversarial Nets (cGAN) and Deep Convolutional Generative Adversarial Networks (DCGAN) in computer vision research [16] [17]. We extended the DCGAN to a conditional version with training data labels and generated the labeled artificial EEG signal.

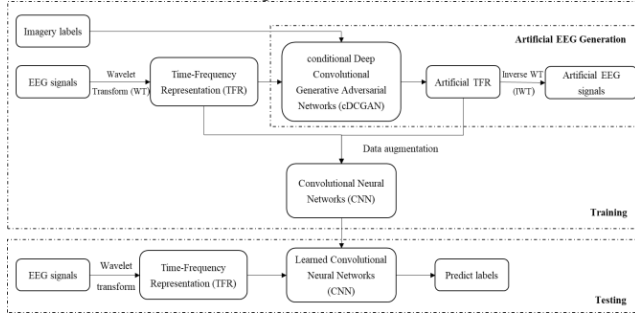

Figure 1. Framework of our proposed data augmentation method

### A. Preprocessing

The patterns that can reflect relevant cognitive activities from EEG signal can be ascribed to several dominant oscillations with specific frequency components, such as theta (4-7Hz), alpha (8-15Hz) and beta (16-30Hz) [18]. Because EEG signal is a kind of non-stationary signal, the wavelet transform is applied to remove noise and extract desired frequency components [19]. In our method, we apply continuous wavelet transform which can lose less information without down sampling and use the complex Morlet wavelet function (CMOR) as basis function [20].

Continuous wavelet transformation of EEG signal $s(t)$ can be defined as,

$$C = \left(\frac{w}{w_0}\right)^{1/2} \int_{-\infty}^{+\infty} s(t')\psi * \left(\frac{w}{w_0}\right)(t'-t)dt' = \langle s(t), \psi(t) \rangle \quad (1)$$

Where $\frac{w}{w_0}$ is the scale factor, $\langle \rangle$ denotes inner product of the signal $s(t)$ and basis wavelet function $\psi(t)$ which is CMOR3-3 in our method.

### B. Conditional deep convolutional generative adversarial networks (cDCGAN)

Generative adversarial networks (GAN) consists of generative model $G$ and discriminative model $D$. The generative model can generate artificial data from noise $Z$ and approximate the feature distribution of training data to fool the discriminative model by adversarial process. Similarly, the discriminative model can learn and revise the feature distribution from training data and adversarial process [11]. Instead of training GAN with multilayer perceptron, CNN is used for constructing GAN structure in both discriminative model and generative model [16].

The whole adversarial process can be described as a two player minimax game with loss function $L$:

$$\min_G \max_D L(D, G) = \mathbb{E}_{x \sim p_{data}(x)}[log D(x)] + \mathbb{E}_{z \sim p_z(z)}[log(1 - D(G(z)))] \quad (2)$$

Where $p_{data}$ is the distribution of training data, $D(x; \theta_d)$ is discriminative model that can estimate the probability that $x$ is from $p_{data}$ by CNN with parameters $\theta_d$, $p_z$ is the distribution of noise and here we use Gaussian distribution, $G(z; \theta_g)$ is generative model that can generate the artificial data from noise $z$ by inverse CNN with parameters $\theta_g$.

To improve the classification accuracy of EEG signal with generated artificial EEG data, the artificial data must have label information. So abovementioned DCGAN is extended to a conditional version with label information, the auxiliary label information $Y$ is feed into both generative model $G$ and discriminative model $D$ [17]. The loss function is converted into a conditional version as follows:

$$\min_G \max_D L(D, G) = \mathbb{E}_{x \sim p_{data}(x)}[log D(x|y_z)] + \mathbb{E}_{z \sim p_z(z)}[log(1 - D(G(z|y_d)))] \quad (3)$$

To avoid overfitting problem on finite dataset during training process, generative model and discriminative model are optimized alternately. In our model, we alternately optimize discriminative model $D$ and generative model $G$ by two steps and one step respectively. Theoretically, as long as $D$ and $G$ are well-designed and trained effectively, the whole adversarial process can reach the Nash equilibrium. Also at this equilibrium, the distribution of artificial data is same as the training data and the artificial data can be seen as real data because the discriminative model cannot distinguish whether it is artificial data or real data. So we use the generative model to generate artificial data from Gaussian noise as extra training data to augment the size of training data.

### A. Convolutional neural networks

To capture enough information from EEG signal in time, frequency and spatial, we applied two-dimensional kernel with multichannel to the preprocessed TFR of EEG signal in CNN model. And our CNN model consists of convolutional layers, pooling layers and fully connected layers.

The TFR of EEG signal $X$ is fed into our CNN model, and $l$-th convolutional layer is applied to the output of the previous layer $X^{l-1}$. After convolutional operation, an additive bias $B_i^l$ is applied and the result is passed through a activation function $f$. Rectified Linear Unit (Relu) is adopted as activation function in our architecture which shows better performance among many activation functions [21]. Denote $W_i^l$ as the convolution kernel in layer $l$ of the network, and let $X^l$ be the TFR of the input at layer $l$. The $i$-th output of convolutional layer $l$ from previous layer is given as follow:

$$X_i^l = f(X_i^{l-1} * W_i^l + B_i^l) X_i^l = f(X_i^{l-1} * W_i^l + B_i^l) \\ = f\left(\sum_{C^l} \sum_{m=1}^{M^l} \sum_{n=1}^{N^l} x_{mn}^{l-1} \times w_{mn}^{l-1} + b_{mn}^l\right) \quad (4)$$

Where $*$ is convolution operation; $M^l$ and $N^l$ are kernel sizes along time axis and frequency axis; $C^l$ is number of kernels in layer $l$; $w_{mn}^l$ is $(m, n)$-th value of 2D kernel in layer $l$; $x_{mn}^{l-1}$ is the $(m, n)$-th value of previous layer input; $b_{mn}^l$ is the $(m, n)$-th value of bias in layer $l$.

## III. MODEL EVALUATION AND RESULTS

To evaluate our method, BCI competition II dataset III is used in this paper. This dataset is about motor imagery task performed by a normal female and EEG signal was recorded over three electrodes (C3, Cz and C4) when subject imagining left or right hand movements. The whole experiment consists of 7 runs with 40 trials and each trial lasts 9 seconds [22]. We used 140 trials of training data to train and optimize our model and 140 trials of testing data to test its performance. As the dataset description instructed, we only extracted 5 seconds length signal between 4s to 9s for further processing and only alpha (7-15Hz) frequency components were extracted for TFR by continuous wavelet transform in this dataset. We mainly studied the quality of generated artificial data and the effect of data augmentation by our proposed cDCGAN method in this preliminary study.

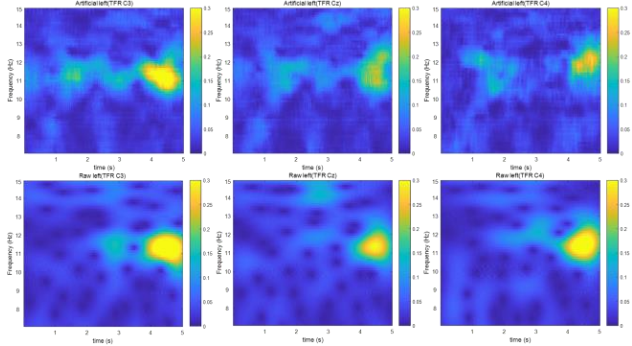

Figure 2. Comparison between artificial EEG time-frequency representation and raw EEG time-frequency representation

At the beginning of training process of our cDCGAN method, the loss of generative model increased fast to a high value but the loss of discriminative model decreased to a very low value which means the ability of discriminative model is much stronger than generative model and it can easily recognize the artificial data in this stage. After thousands of iteration epochs, the loss of generative model and discriminative model changed alternately and the accuracy of discriminative model converged gradually to 50% with oscillation which means the two models in an adversarial process and struggle for approximating to an equilibrium. Theoretically, in this equilibrium the distribution of artificial data is same as the training data and the artificial data can be seen as real data because the discriminative model cannot distinguish whether it is artificial data or real data [11]. In Fig. 2, we showed one of our generated artificial EEG TFR in C3, Cz and C4 when the subject imagined left hand movements. Compared with the TFR of raw EEG signal, the generated artificial TFR had the consistent principal time-frequency features in all three channels with raw EEG signal and had other additional features. Also all the features of artificial EEG data are learned from the feature distribution of raw EEG data instead of just remembering some of the raw data.

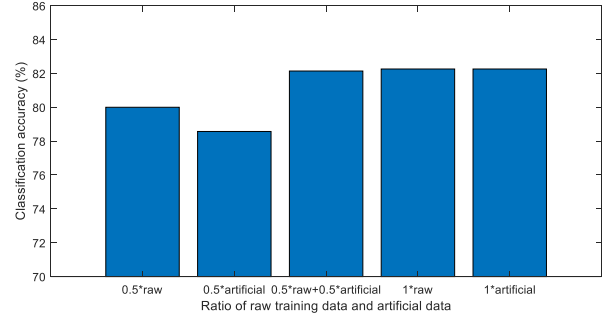

Figure 3. Classification accuracy of testing dataset for different ratio of raw training data and artificial data.

To evaluate the quality of our generated artificial EEG data, we trained the same CNN model with 5 groups of different training data and tested the classification performance of trained model in the same testing dataset. We took the size of raw training data (70 samples for per class) as reference and five groups of training data were designed by different ratio of training data and artificial data. As Fig. 3. showed, the classification accuracy of raw EEG training data (1*raw), artificial EEG data (1*artificial) and mixed EEG data (0.5*raw+0.5*artificial) had the nearly same classification accuracy with 82.86%, 82.86 and 82.14% respectively. This results verified the quality of our generated artificial EEG data by our cDCGAN method that has no less than raw EEG data, and CNN model with the same structure can learn enough features from our generated artificial data as much as the raw data does. Classification performance of both raw data and artificial data suffered from lack of data, when we only used half of the data classification accuracy of both raw data and artificial data decreased.

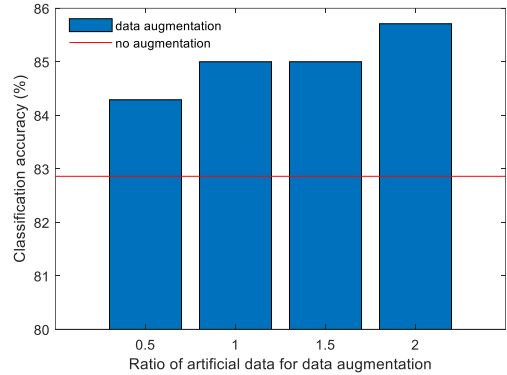

Figure 4. Classification accuracy of testing dataset for different ratio of artificial data for data augmentation.

To evaluate the effect of our proposed data augmentation method, we trained the same CNN model using mixed training data with different ratio of artificial data for data augmentation. We took the size of raw training data (70 samples for per class) as reference and mixed 0.5, 1, 1.5 and 2 times artificial data with raw training data for data augmentation which means added 35, 70, 105 and 140 artificial samples per class. As Fig. 4. showed our proposed data augmentation method can effectively improve the classification accuracy in testing dataset than the raw training data without data augmentation. And with the number of artificial data increasing, the classification performance of data augmentation increases.

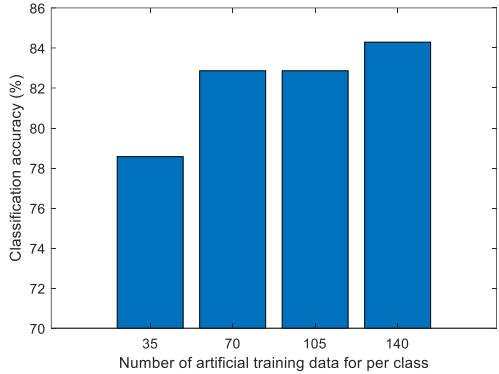

Figure 5. Classification accuracy of testing dataset for different number of artificial data per class.

We also studied the effect of different number of training data to the classification performance using artificial data. As Fig. 5. showed the number of artificial training data can significantly influence classification performance in testing data. When the number of training data per class was less than 70, the improvements were more obvious. This results were also consistent with our prior studies, in that study we reassigned the ratio of raw training and testing dataset (RTT) with 1:4, 1:3, 1:2, 1:1, 2:1, 3:1 and 4:1 seven different situations and the classification accuracy of all five subjects increased with increasing of RTT, also the improvements were more obvious when RTT was less than 1 [23].

IV. CONCLUSION AND FUTURE WORK

With the development of brain computer interface, many algorithms are proposed to improve its performance. But data scarcity is a big problem confines the ability of these advanced algorithms especially for the deep learning methods, because limited data cannot exploit the full potential of deep learning model in BCI field. So, we proposed a novel approach to generate more artificial brain signal data automatically to overcome this data deficiency problem and improve the performance of deep learning model. The quality of our artificial EEG data and the performance of our data augmentation method have been evaluated in BCI competition dataset, it can effectively improve classification accuracy with our cDCGAN method for data augmentation. Although our proposed method is very promising, we only evaluate its classification performance in CNN model. In the future, we will study the generated artificial EEG data in more intrinsic and interpretable perspective.